\documentclass[a4paper]{spie}  
\usepackage[]{graphicx}

\title{The QACITS pointing sensor: from theory to on-sky operation on Keck/NIRC2}

\author{
Elsa\,Huby\supit{a}, 
Olivier\,Absil\supit{a}, 
Dimitri\,Mawet\supit{b}, 
Pierre\,Baudoz\supit{c}, 
Bruno\,Femen\`ia Castell\`a\supit{d}, 
Michael\,Bottom\supit{b}, 
Henry\,Ngo\supit{b} and
Eugene\,Serabyn\supit{e}
\skiplinehalf
\supit{a}Space sciences, Technologies and Astrophysics Research (STAR) Institute, Universit\'e de Li\`ege, 19c All\'ee du Six Ao\^ut, B-4000 Li\`ege, Belgium; \\
\supit{b}California Institute of Technology, 1200 E. California Blvd, Pasadena, CA 91125, USA;\\
\supit{c}LESIA, Observatoire de Paris, CNRS, UPMC, Universit\'e Paris-Diderot, Paris Sciences et Lettres, 5 place Jules Janssen, 92195 Meudon, France;\\
\supit{d}W. M. Keck Observatory, 65-1120 Mamalahoa Hwy., Kamuela, HI 96743, USA;\\
\supit{e}Jet Propulsion Laboratory, 4800 Oak Grove Dr., Pasadena, CA 91109, USA
}

\authorinfo{Send e-mail correspondence to elsa.huby@ulg.ac.be}

\begin{document} 
  \maketitle 

\begin{abstract}
Small inner working angle coronagraphs are essential to benefit from the full potential of large and future extremely large ground-based telescopes, especially in the context of the detection and characterization of exoplanets. Among existing solutions, the vortex coronagraph stands as one of the most effective and promising solutions. However, for focal-plane coronagraph, a small inner working angle comes necessarily at the cost of a high sensitivity to pointing errors. This is the reason why a pointing control system is imperative to stabilize the star on the vortex center against pointing drifts due to mechanical flexures, that generally occur during observation due for instance to temperature and/or gravity variations. We have therefore developed a technique called QACITS\cite{Huby2015} (Quadrant Analysis of Coronagraphic Images for Tip-tilt Sensing), which is based on the analysis of the coronagraphic image shape to infer the amount of pointing error. It has been shown that the flux gradient in the image is directly related to the amount of tip-tilt affecting the beam. The main advantage of this technique is that it does not require any additional setup and can thus be easily implemented on all current facilities equipped with a vortex phase mask. In this paper, we focus on the implementation of the QACITS sensor at Keck/NIRC2, where an L-band AGPM has been recently commissioned (June and October 2015), successfully validating the QACITS estimator in the case of a centrally obstructed pupil. The algorithm has been designed to be easily handled by any user observing in vortex mode, which is available for science in shared risk mode since 2016B.
\end{abstract}



\section{INTRODUCTION}
\label{sec:intro} 

Small inner working angle (IWA) coronagraphs are the key to access the full angular resolution potential of current large ground based telescopes. The IWA is defined as the angular separation at which the flux of an off-axis companion is transmitted by 50\%. Among the existing solutions, the vector vortex coronagraph is based on a focal plane phase mask inducing a phase ramp onto the beam. This kind of coronagraph is attractive for several reasons, including high extinction ratio, achromatic properties, a continuous discovery space and small IWA. For all these reasons, this kind of coronagraph can be found in several leading instruments: the Palomar infrared camera PHARO \cite{Mawet2010a}, Subaru/SCExAO \cite{Jovanovic2015}, VLT/VISIR \cite{Delacroix2012,Kerber2014}, VLT/NACO\cite{Mawet2013}, LBT/LMIRCam \cite{Defrere2014}  and recently Keck/NIRC2 (first light obtained in June 2015, see Ref.~\citenum{FemeniaCastella2016} in these proceedings and Serabyn et al. in prep). The latter four are mid-infrared instruments working in the L or N band and are equipped with Annular Groove Phase Masks\cite{Mawet2005} (AGPM). These components are developed by the University of Li\`ege and manufactured by the University of Uppsala\cite{Forsberg2014}. For a review of the results obtained with these instruments, see Ref.~\citenum{Absil2016} in these proceedings.

\begin{figure}
\begin{center}
\begin{tabular}{c}
\includegraphics[height=6cm]{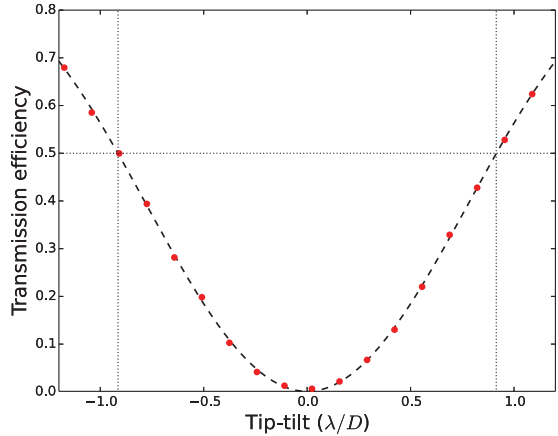}
\end{tabular}
\end{center}
\caption[transmission]
{  \label{fig:transmission}
Experimental transmission of a vortex phase mask (AGPM of topological charge 2) for a circular non obstructed pupil. The flux is integrated on a disk of diameter equal to the full width at half maximum. Estimated IWA is 0.9\,$\lambda/D$ as predicted by simulations.}
\end{figure}

However, a small IWA focal-plane coronagraph is also a synonym for high sensitivity to pointing error. Indeed, a slight shift of the star from the mask center will result in a starlight leakage and degrade the contrast performance. A low order wavefront sensor is therefore required in high contrast imaging instruments (for a full review of low order wavefront sensor possibilities, see Ref.~\citenum{Mawet2012}), in addition to the Adaptive Optics system. Fig.~\ref{fig:transmission} shows the experimental transmission measured in L band with an AGPM. This curve is valid for an off-axis companion as well as for the central star. In order to estimate the non common path aberrations, this additional sensor has to be placed as close as possible to the coronagraphic mask and/or science camera. In the case of the vortex phase mask, we have developed a pointing sensor algorithm called QACITS\cite{Huby2015} (Quadrant Analysis of Coronagraphic Images for Tip-tilt Sensing). While the principle has first been empirically introduced and implemented in laboratory for the Four Quadrant Phase Mask\cite{Mas2012}, we have derived the complete theoretical framework adapted to the vortex coronagraph\cite{Huby2015} and successfully implemented it on-sky.


\section{THE QACITS PRINCIPLE}
\label{sec:principle}

The principle of QACITS relies on a flux measurement in the image, and in particular on the flux difference along two orthogonal axes, thus measuring the amplitude of the flux asymmetry. 

\subsection{Analytical derivation for a circular pupil}

The complete analytical derivation is described in detail in Ref.~\citenum{Huby2015} and will not be repeated here. It is based on two assumptions: (i) the pupil is circular and non obstructed, such that the wavefront can be described using Zernike polynomials; and (ii) the amplitude of the aberrations is small, allowing to approximate the expression of the wavefront by the complex phase term (first order Taylor expansion of the exponential term).

The main results of this analytical modelling are the following:
\begin{itemize}
\item if the wavefront is described by a Zernike mode at the entrance of the coronagraph (with a small amplitude), the wavefront defined at the Lyot plane is discontinuous and consists of two components that are defined inside and/or outside the geometrical pupil (see Fig.\,A.1 in Ref.\,\citenum{Huby2015});
\item the part of the wavefront that is defined inside the geometrical pupil (i.e. transmitted through the coronagraph) can be written as a complex linear combination of Zernike polynomials (see Table A.1 in Ref.\,\citenum{Huby2015} for a conversion table);
\item when describing the wavefront using the second order exponential approximation, the asymmetry of the final image appears through the term (all the other terms produce a symmetric pattern): 
\begin{equation}
I_{\rm det}^{\rm asym} = 2 T_x^3 \cos \theta \frac{2 J_2(\alpha)}{\alpha} \frac{2 J_3(\alpha)}{\alpha}
\end{equation}
where $T_x$ is the amplitude of the tip-tilt (considered in the $x$ direction only), $(\alpha, \theta)$ are the polar coordinates in the image plane and $J_2$ and $J_3$ are the Bessel functions of the first kind. As a result, the differential intensity is proportional to the cube of the tip-tilt amplitude:
\begin{equation}
\Delta I _ x  = \int _0^{\infty } \int _ {- \frac{\pi}{2}} ^ {\frac{\pi}{2}} I_{\rm det}^{\rm asym} (\alpha, \theta) {\rm d}\theta{\rm d}\alpha  - \int _0^{\infty } \int _ {\frac{\pi}{2}} ^ {\frac{3\pi}{2}} I_{\rm det}^{\rm asym} (\alpha, \theta) {\rm d}\theta{\rm d}\alpha \propto T_x^3
\end{equation}
\end{itemize}

\subsection{The case of a centrally obstructed pupil}

\begin{figure}
\begin{center}
\begin{tabular}{rc}
\textbf{a)} & 25\% central obstruction (Keck) \\
 & \includegraphics[width=15cm]{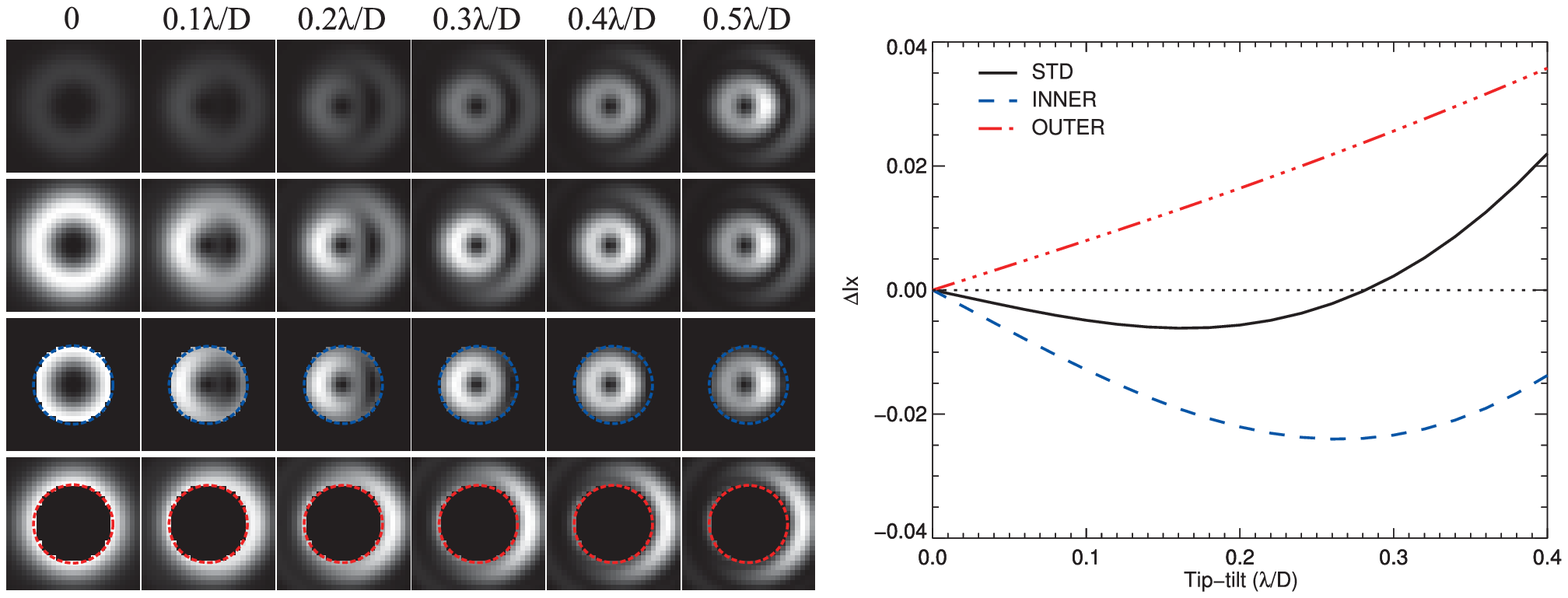} \\
\textbf{b)} & 14\% central obstruction (VLT) \\
 & \includegraphics[width=15cm]{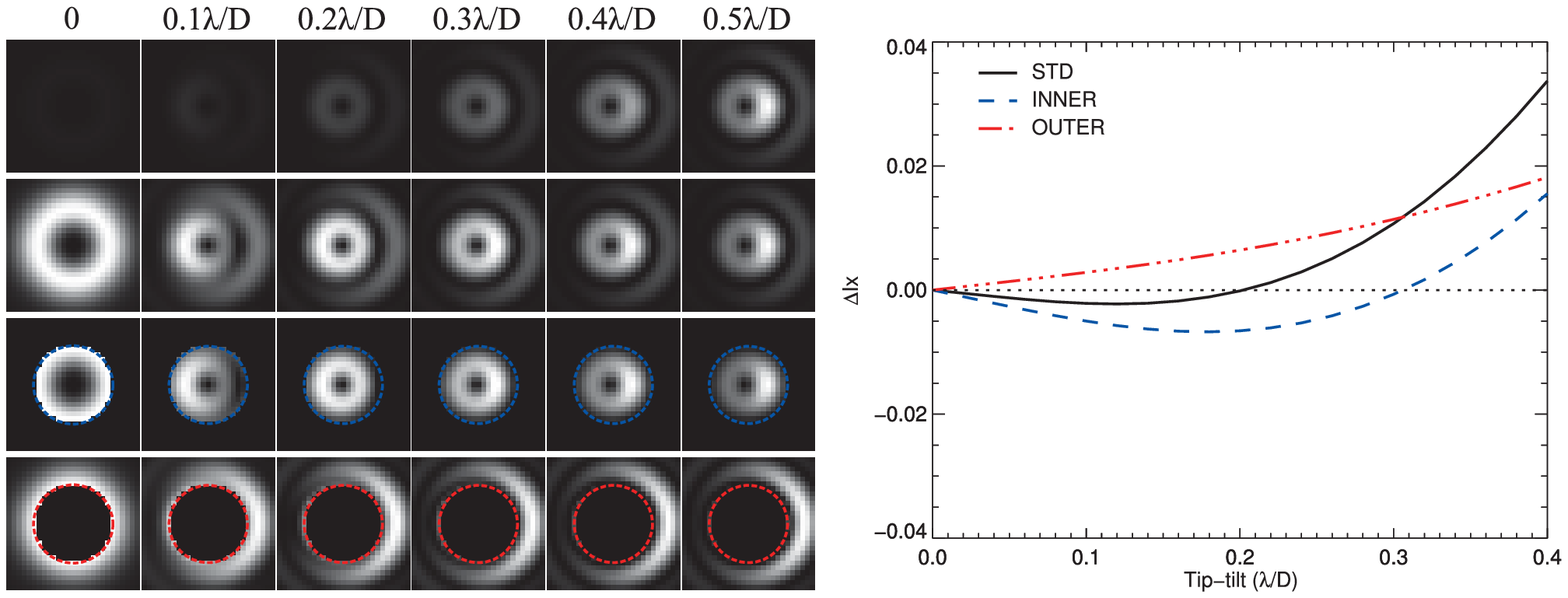}
\end{tabular}
\end{center}
\caption[qacits_simu]{
\label{fig:qacits_simu}
Simulation of coronagraphic images for different amplitudes of tip-tilt (given at the top of the image rows) for \textbf{a)} a Keck-like pupil (25\% central obstruction) and \textbf{b)} a VLT-like pupil (14\% central obstruction). In each case, the first and second rows of images show the whole images with a constant and stretched color bar, respectively. The bottom two rows of images highlight the central and the external parts used in the inner and outer QACITS estimator. The graphs on the right-hand side show the the resulting differential intensity measurement as a function of tip-tilt, when taking the whole image into account (STD) or only the inner (INNER) or the outer (OUTER) part of the image.
}
\end{figure}

The analytical derivation can also be carried out in the case of a centrally obstructed pupil, by decomposing the wavefront into a positive component of diameter $D$ (external diameter) and a negative component of diameter $d<D$ (central obstruction). In Ref.\,\citenum{Huby2015}, it has been shown that in the presence of a central obstruction, the final image on the detector includes a term that is directly proportional to the cube of the amount of tip-tilt (contribution of the circular pupil), and another term that is directly proportional to the tip-tilt amplitude (contribution of the central obstruction). Both terms compensate for each other, making the flux intensity difference not monotonic if computed on the whole image, as shown in Fig.\,\ref{fig:qacits_simu}. However, the disentanglement of these two contributions is possible when considering the inner and the outer part of the image separately. The limit between the two areas is chosen at 1.6$\lambda/{D_{Lyot}}$, where the Bessel functions change sign (see Fig.~6 in Ref.~\citenum{Huby2015}). In this case, the differential flux is monotonic to some extend and the tip-tilt amplitude can be retrieved. As illustrated in Fig.\,\ref{fig:qacits_simu}, the coefficients of the linear model depends on the size of the central obstruction.

\section{IMPLEMENTATION AT KECK/NIRC2}
\label{sec:implementation}

An L band AGPM has recently been integrated in the Keck/NIRC2 instrument (see Ref~\citenum{FemeniaCastella2016} in these proceedings for a detailed description of the integration, and Serabyn et al. in prep). First light of this new mode was obtained on June, 8th, 2015. On the second night of this commissioning run, the QACITS loop was successfully closed. Another commissioning run took place in October 2015, during which the current QACITS script has been developed and tested on-sky. In practice, only the outer QACITS estimator is used (the inner part of the image is likely affected by residual speckles that bias the inner QACITS estimator) and the loop is based on a proportional-integral controller with typical gain of 0.9 and 0.3, respectively.

\subsection{The QACITS procedure}

The script takes care of every calibration steps as well as science acquisitions. The whole QACITS procedure consists of three automated steps:

\begin{description}
\item[Calibration step] An off-axis PSF image and its associated sky image are acquired (both are needed to calibrate the flux in the QACITS estimator as well as for the data processing when a PSF template is required). A sky image associated with the science/optimization images is also acquired.
\item[Optimization step] The star is centered onto the vortex center using the QACITS estimator. Acquisition settings are chosen to allow faster acquisitions (reduced integration time and number of co-added frames, as well as sub-framed images). This loop stops as soon as the tip-tilt estimations are stable (by default, there should be two consecutive estimations smaller than 0.1$\lambda/D$, but these criterion can be tuned by the observer). If this criterion is not met, the loop will stop after a maximum of 10 iterations.
\item[Science acquisition sequence] The QACITS algorithm is run in closed loop while acquiring data with the desired acquisition settings for science images.
\end{description}

The different steps listed above can be run separately or all in a row (default use).

\subsection{Parameters and variables}

There are several parameters that should be given by the observer as inputs of the IDL calling sequence. Typical parameters that can be defined as inputs of the QACITS calling sequence are listed in Table \ref{tab:inputs}. The variables defining the path of the data directory and the science acquisition settings are required, while some others are optional and can be used as keywords.

\begin{table}[h]
\caption{Typical inputs of the IDL calling standard calling sequence.}
\label{tab:inputs}
\begin{center}
\begin{tabular}{|l|l|l|}
\hline
\rule[-1ex]{0pt}{3.5ex}  Variable name &  Type & Description  \\
\hline
\rule[-1ex]{0pt}{3.5ex}  \verb|n_sci| & \textit{required} & Number of acquisitions desired for the science sequence  \\
\hline
\rule[-1ex]{0pt}{3.5ex}  \verb|tint_sci| & \textit{required} & Integration time for the science acquisition frames  \\
\hline
\rule[-1ex]{0pt}{3.5ex}  \verb|coad_sci| & \textit{required} & Number of co-added frames for the science acquisitions  \\
\hline
\rule[-1ex]{0pt}{3.5ex}  \verb|data_dir| & \textit{required} & Path name of the directory where the data are saved (string)  \\
\hline
\rule[-1ex]{0pt}{3.5ex}  \verb|tint_opti| & \textit{optional} & Integration time for the optimization acquisitions  \\
\hline
\rule[-1ex]{0pt}{3.5ex}  \verb|coad_opti| & \textit{optional} & Number of co-added frames for the optimization acquisitions  \\
\hline
\rule[-1ex]{0pt}{3.5ex}  \verb|subc_opti| & \textit{optional} & Width of the image during optimization (typically 1024, 512, 256 or 128)  \\
\hline
\rule[-1ex]{0pt}{3.5ex}  \verb|/no_calib| & \textit{optional} & For skipping the calibration sequence  \\
\hline
\rule[-1ex]{0pt}{3.5ex}  \verb|/no_opti| & \textit{optional} & For skipping the optimization sequence  \\
\hline
\rule[-1ex]{0pt}{3.5ex}  \verb|/faint| & \textit{optional} & Faint target mode (optimization settings set to science acquisition settings)  \\
\hline
\end{tabular}
\end{center}
\end{table}

There are other parameters that can be tuned by the observer if necessary, but they are set to default values that should work in most cases. The observer can only access these values by modifying a QACITS script called \verb|qacits_nirc2_params.pro|. These parameters, along with their default values, are listed in Table~\ref{tab:params}.

\begin{table}[h]
\caption{Advanced parameter settings.}
\label{tab:params}
\begin{center}
\begin{tabular}{|l|l|l|}
\hline
\rule[-1ex]{0pt}{3.5ex}  Variable name & Default value & Description  \\
\hline
\rule[-1ex]{0pt}{3.5ex}  \verb|tint_opti| & 0.2\,s & Integration time for the optimization acquisitions \\
\hline
\rule[-1ex]{0pt}{3.5ex}  \verb|coad_opti| & 10 & Number of co-added frames for the optimization acquisitions  \\
\hline
\rule[-1ex]{0pt}{3.5ex}  \verb|subc_opti| & 512 & Width of the image during optimization  \\
\hline
\rule[-1ex]{0pt}{3.5ex}  \verb|tint_psf| & 0 &  Off-axis PSF integration time \\
\hline
\rule[-1ex]{0pt}{3.5ex}  \verb|coad_psf| & 100 & Number of co-added frames for the off-axis PSF  \\
\hline
\rule[-1ex]{0pt}{3.5ex}  \verb|subc_psf| & 128 & Width of the image for the off-axis PSF  \\
\hline
\rule[-1ex]{0pt}{3.5ex}  \verb|faint_tint_psf| & 0.2\,s &Integration time for the off-axis PSF in the /faint mode  \\
\hline
\end{tabular}
\end{center}
\end{table}

\subsection{Calling sequences}

QACITS is usually used to perform the calibration, centering optimization and science acquisition automatically. However, every step can be run separately if needed, but this it usually not required nor recommended, except for extra science acquisition sequences. Science acquisition sequences can be launched either using specific keywords to skip the calibration and optimization steps in the standard calling sequence (see Section \ref{sec:std_seq}) or by using the dedicated calling sequence (see Section \ref{sec:sci_seq}).

\subsubsection{Standard sequence}
\label{sec:std_seq}

By default, this sequence includes the three steps: calibration, optimization and science acquisitions. The IDL calling sequence is the following:\\
\verb|IDL> run_qacits_nirc2, n_sci, tint_sci, coad_sci, data_dir=data_dir| \\

If you want to skip both calibration and optimization steps, you can use the \verb|/no_calib| and/or \verb|/no_opti| keywords (it is then equivalent to running a science sequence alone as described in Section \ref{sec:sci_seq}):\\
\verb|IDL> run_qacits_nirc2, n_sci, tint_sci, coad_sci, data_dir=data_dir, /no_calib, /no_opti|\\

The acquisition settings for the optimization sequence have default values (see Table~\ref{tab:params}), but they can be set to custom values by using the optional keywords:\\
\verb|IDL> run_qacits_nirc2, n_sci, tint_sci, coad_sci, data_dir=data_dir, tint_opti=tint_opti,| \\
\verb|coad_opti=coad_opti, subc_opti=subc_opti|\\

\paragraph{Faint mode} There is a \verb|/faint| keyword that can be used for faint targets ($L_{\rm mag} > 5$). In the \verb|/faint| mode, the integration time for the off-axis PSF is set to a default 0.2 s value (see Table~\ref{tab:params}) to get a reasonable SNR. Besides, the acquisition settings for the optimization sequence are set to the same as for the science sequence. In case of faint targets, low integration time and number of co-added frames are not appropriate even for the optimization sequence. Therefore the acquisition settings are set to the science settings, such that these acquisitions can potentially be used as science frames, but are still flagged as optimization frames.

\subsubsection{Calibration sequence}
\label{sec:calib_seq}

The calibration sequence can be run separately using: \\
\verb|IDL> run_qacits_calib, tint_sci, coad_sci, data_dir=data_dir|\\
Note that the star should not be too far from the vortex center at the beginning of the sequence, because the star needs to be in the field of view after an offset by $\sim 1000$\,pixels from the initial position. At the end of the calibration sequence, the star will be roughly centered onto the vortex. Indeed, the position of the vortex center is estimated from the sky image, and the position off-axis PSF can be fitted as well, allowing to estimate the offset and correct for it.

\begin{figure}
\begin{center}
\begin{tabular}{c}
\includegraphics[scale=0.4, angle=270]{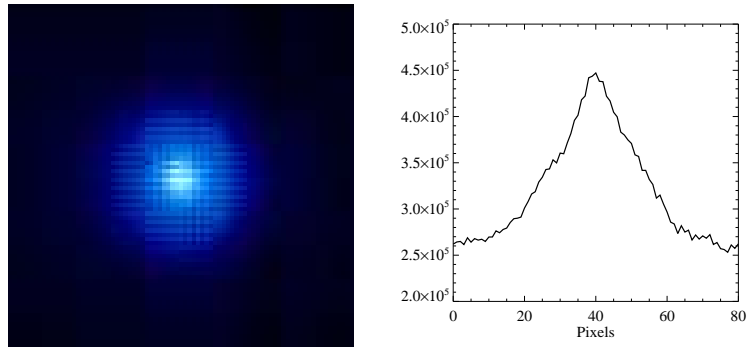}
\end{tabular}
\end{center}
\caption[center_emission]
{  \label{fig:emission}
\textit{Left:} On-sky image of the vortex center without any star centered. A thermal signal emitted by the center of the vortex is observed. \textit{Right:} Horizontal profile of this signal.}
\end{figure}

\paragraph{Estimation of the vortex mask center position}
As described in more details in Ref.~\citenum{FemeniaCastella2016} and \citenum{Absil2016} in these proceedings, a signal emitted by the vortex center is observed in the sky images (while the star is offset in the field). This particular signal is shown in Fig.~\ref{fig:emission}. In the QACITS algorithm, this pattern is used to precisely estimate the center of the vortex by fitting a 2D Gaussian function. The vortex center position is used at the end of the calibration sequence to roughly recenter the star after the acquisition of the off-axis PSF image, and it is also used later in the QACITS algorithm as the reference position of the vortex center.

\subsubsection{Optimization sequence}
\label{sec:opti_seq}

The optimization sequence for finely centering the star onto the vortex mask can be run by calling:\\
\verb|IDL> run_qacits_opti, tint_sci, coad_sci, data_dir=data_dir|\\
For a maximized efficiency, the star should be roughly centered onto the vortex center (the image should at least exhibit a donut shape). Note that \verb|tint_sci| and \verb|coad_sci| keywords should be provided in case you want to use the \verb|/faint| mode.
The optimization sequence is acquired with default values but there are additional optional keywords that can be used by the observer:\\
\verb|IDL> run_qacits_opti, tint_sci, coad_sci, data_dir=data_dir,tint_opti=tint_opti,|\\
\verb|coad_opti=coad_opti, /do_calib|\\
The keyword \verb|/do_calib| will force the calibration sequence to be called. If \verb|/do_calib| is not set and if calibration files already exist for this target, they will be used and the calibration sequence will be
automatically skipped. If they do not exist, the calibration sequence will be run automatically. A calibration sequence should better be performed again if observations on a target are longer than 1h, to refresh the sky image and off-axis PSF, which are likely to vary with time.

\subsubsection{Science acquisition sequence}
\label{sec:sci_seq}

The science acquisition sequence can be run using the dedicated calling sequence:\\
\verb|IDL> run_qacits_sci, n_sci, tint_sci, coad_sci, data_dir=data_dir|

\subsection{The log file}
A log file is automatically generated when acquisitions are taken by the QACITS routines (note that the acquisitions taken using the \verb|goi| command in the nirc2 terminal are not logged in this file). This file is
named as \verb|[UT-date]_nirc2_qacits.log| and can be found in the folder called log in the QACITS folder. The observer has to make sure that this log folder exists when starting the observation. The different columns, delimited by a semi-colon are:
\begin{itemize}\itemsep1pt
\item the acquisition number of the file;
\item the UT time;
\item the name of the object;
\item the acquisition type (see Table \ref{tab:type});
\item the integration time;
\item the number of co-added frames;
\item the dtclxoffset status (if relevant);
\item the dtclyoffset status (if relevant);
\item the estimated tip-tilt in x (if relevant);
\item the estimated tip-tilt in y (if relevant);
\item the estimation of the null depth (if relevant).
\end{itemize}

\begin{table}[h]
\caption{Acquisition types}
\label{tab:type}
\begin{center}
\begin{tabular}{|l|l|}
\hline
\rule[-1ex]{0pt}{3.5ex} Type & Description  \\
\hline
\rule[-1ex]{0pt}{3.5ex} \verb|sci| & Science image  \\
\hline
\rule[-1ex]{0pt}{3.5ex} \verb|sky| & Sky image for the science and optimization images \\
\hline
\rule[-1ex]{0pt}{3.5ex} \verb|psf| & Off-axis PSF image  \\
\hline
\rule[-1ex]{0pt}{3.5ex} \verb|skp| & Sky image for the off-axis PSF  \\
\hline
\rule[-1ex]{0pt}{3.5ex} \verb|opt| & Image taken during the optimization sequence  \\
\hline
\end{tabular}
\end{center}
\end{table}

Typical lines in the log file looks like the following:\\
\fontsize{8}{5}
\verb|num ; time ; object ; acq type ; int time ; coadds ; offx ; offy ; ttx ; tty ; null|\\
\verb|...|\\
\verb|31 ; 06:35:42 ; HR 8799 ; skp ; 0.0180000 ; 100.000 ; 0.0000000 ; 0.0000000 ; nc ; nc ; 0| \\
\verb|32 ; 06:36:10 ; HR 8799 ; psf ; 0.0180000 ; 100.000 ; 0.0000000 ; 0.0000000 ; nc ; nc ; 0| \\
\verb|33 ; 06:37:21 ; HR 8799 ; sky ; 0.181000 ; 100.000 ; 0.0000000 ; 0.0000000 ; nc ; nc ; 0| \\
\verb|34 ; 06:39:58 ; HR 8799 ; opt ; 0.200000 ; 10.0000 ; 0.0000000 ; 0.0000000 ; 0.0623256 ; 0.187051 ; 4.96083| \\
\verb|35 ; 06:40:28 ; HR 8799 ; opt ; 0.200000 ; 10.0000 ; ­0.0098141832 ; ­0.0038910112 ; 0.0649361 ; 0.270143 ; 5.85587| \\
\verb|36 ; 06:40:50 ; HR 8799 ; opt ; 0.200000 ; 10.0000 ; ­0.023988070 ; ­0.0079449946 ; 0.00857234 ; 0.101631 ; 6.94394| \\
\verb|37 ; 06:41:12 ; HR 8799 ; opt ; 0.200000 ; 10.0000 ; ­0.029320449 ; ­0.0079449946 ; 0.00640033 ; ­0.0137577 ; 6.35239| \\
\verb|38 ; 06:43:06 ; HR 8799 ; sci ; 0.500000 ; 50.0000 ; ­0.029200000 ; ­0.0051000000 ; 0.0352203 ; 0.0612146 ; 10.232| \\
\normalsize

\subsection{Displayed windows}

\begin{figure}
\begin{center}
\begin{tabular}{c}
\includegraphics[width=17cm]{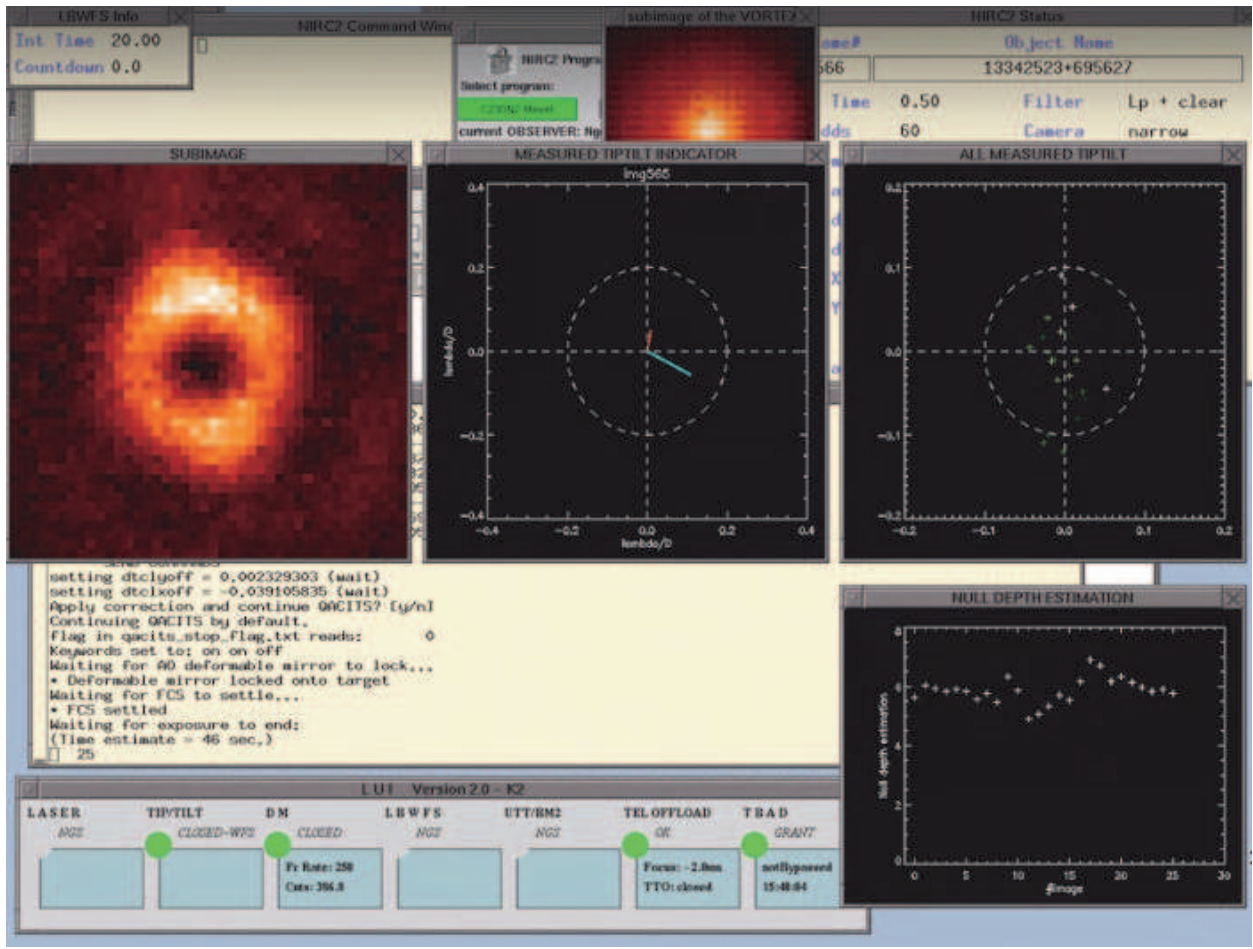}
\end{tabular}
\end{center}
\caption{
\label{fig:display}
Typical display windows appearing when running the QACITS loop.
}
\end{figure}

At the beginning of a sequence, two small display windows will appear at the top of the screen showing the off-axis PSF and the pattern emitted by the center of the vortex mask. If either of this window looks empty or weird, the sequence will most probably crash, and it will be necessary to recenter the star on the vortex manually before starting the sequence again. During the QACITS loop (optimization or science), several windows will appear and be refreshed at every iteration (see Fig.~\ref{fig:display}):
\begin{description}
\item[left:] the sub-image (sky subtracted) used by the QACITS estimator. The inner circle (radius of 2\,$\lambda/D$) of the image is enhanced by 30\%.
\item[middle:] the current estimation of tip-tilt. There are 2 different estimators: the blue vector corresponds to the estimator based on the inner part of the image, while the red one corresponds to the estimator based on the outer part of the image. The green dashed vector is the one that is currently used by QACITS (in practice, only the outer estimator is used).
\item[right:] all estimations of tip-tilt since the beginning of the sequence. The current one is drawn in white, while previous iterations appear in darker shades of green.
\item[bottom right:] rough estimation of the null depth (ratio of the flux integrated over the central disk (radius of 2\,$\lambda/D$) of the coronagraphic image and off-axis PSF). This is meant to monitor the stability of the loop, and does not provide a contrast estimate.
\end{description}

\section{CONCLUSION AND PROSPECTS}

The QACITS algorithm has been specifically developed for stabilizing the positioning of a star onto the center of a vortex coronagraph. This algorithm is necessary to optimize the extinction performance of the coronagraph. In addition, the procedure that we have implemented at Keck/NIRC2 is fully automated and takes care of every steps of the acquisition: calibration, centering optimization, science frames acquisition. As a result, the vortex observation mode is now "user-friendly" and optimized in efficiency (manual alignment can be much longer and tiresome for the observer). Besides, the ease of operation allowed by the QACITS procedure has made possible to offer the vortex mode for science in shared risk mode since 2016B.

The simplicity of its implementation and its robustness make QACITS very attractive for other existing vortex modes. Preliminary tests have been carried out at LBT/LMIRCam (in collaboration with D.\,Defr\`ere from Universit\'e de Li\`ege), and at VLT/NACO (in collaboration with J.\,Girard and G.\,Zins from ESO). The results are very promising, and the implementation of a NACO template dedicated to observation with the vortex mode is currently under consideration.

\bibliography{VORTEXbib}   
\bibliographystyle{spiebib}   

\end{document}